\documentclass[a4paper,runningheads]{llncs}

\newcommand*{\coanote}[1]{}

\usepackage{enumerate}
\usepackage{ifthen}
\usepackage{hevea}
\usepackage{url}

\newboolean{abstractonly}
\setboolean{abstractonly}{false}

\ifthenelse{\boolean{hevea}}{
\newcommand*{\Cplusplus}{C++}
}{
\newcommand*{\Cplusplus}{{C\nolinebreak[4]\hspace{-.05em}\raisebox{.4ex}{\tiny\bf ++}}}
}

\begin{document}

\title{%
  A Prolog-based Environment \\
  for Reasoning about Programming Languages \\
  (Extended abstract)
}

\ifthenelse{\boolean{hevea}}{
\author{Roberto Bagnara,
    Department of Mathematics,
    University of Parma,
    Italy,
    \url{bagnara@cs.unipr.it} \\
\and Patricia M. Hill,
    School of Computing,
    University of Leeds,
    UK,
    \url{hill@comp.leeds.ac.uk} \\
\and Enea Zaffanella,
    Department of Mathematics,
    University of Parma,
    Italy,
    \url{zaffanella@cs.unipr.it}
}
}{
\author{Roberto Bagnara\inst{1}
\and Patricia M. Hill\inst{2}
\and Enea Zaffanella\inst{1}
}
\authorrunning{R.~Bagnara,
               P.~M.~Hill,
               E.~Zaffanella
}
\titlerunning{%
  A Prolog-based environment for reasoning about programming languages
}

\institute{
    Department of Mathematics,
    University of Parma,
    Italy \\
    \email{\{bagnara,%
             zaffanella\}@cs.unipr.it}
\and
    School of Computing,
    University of Leeds,
    UK \\
    \email{hill@comp.leeds.ac.uk}
}
}

\maketitle
\ifthenelse{\boolean{abstractonly}}{
\begin{abstract}
}{
}
\emph{ECLAIR} is a Prolog-based prototype system aiming to provide
a functionally complete environment for the study, development
and evaluation of programming language analysis and implementation tools.
In this paper, we sketch the overall structure of the system, outlining the
main methodologies and technologies underlying its components.
We also discuss the appropriateness of Prolog as the implementation
language for the system: besides highlighting its strengths,
we also point out a few potential weaknesses,
hinting at possible solutions.
\ifthenelse{\boolean{abstractonly}}{
\end{abstract}
\end{document}
}{}

\subsubsection*{Motivation for a flexible language resource.}
Static program analysis aims to find bugs, verify the absence of
errors in software, and ensure the correctness of given optimizations.
The standard and most used theoretical framework
that lies behind static program analysis is abstract interpretation
\cite{CousotC77,CousotC92fr}. This framework, which has been available
for more than thirty years, allows us to separate the \emph{abstract
domains}, for representing the program's properties of interest, from
the \emph{abstract interpreter} that should mimic the execution of the
programs on these domains.  As far as the abstract domains are
concerned, there is now a good choice of
implementations offering a flexible precision/efficiency trade-off.
For the abstract interpreter, one non-commercial interpreter
that has been used for automatically verifying
up to one million lines of C code is \emph{ASTR\'EE} \cite{CousotCFMMMR05};
however, ASTR\'EE is specially targeted at a particular class of
programs and program properties, so that widening its scope of
application is likely to require significant effort \cite{Cousot05b}.
The interpreters provided by ECLAIR system follow the methodology described
in~\cite{BagnaraHPZ07TR} which handles most of the problematic
features of mainstream, single-threaded languages;
the concrete semantics is based on the structured operational semantics
extended to allow for infinite computations \cite{CousotC92,Kahn87,Plotkin04b}
and the abstract semantics is based on the work of
Schmidt~\cite{Schmidt95,Schmidt97,Schmidt98}.
The ECLAIR system is being designed to have a
high degree of flexibility for the following reasons:
\begin{itemize}
\item
As there is a constantly evolving and, potentially, huge set of
programming languages for which analyzers are needed, the prototype
interpreter needs to target a variety of language features (including,
for example, exception handling, functions and pointers) and have the
capability of being extended to incorporate further languages and
language features, not yet considered.  Thus, the overall environment
has to be designed to provide a uniform interface that
delegates the details of the implementation of the run time system to
independent components.
\item
In order to analyze for different types of program properties and
error states, the system has to be able to connect with a wide variety
of abstract domains.  Note that the system's interface with the
abstract domains and their operations must be designed so that it can
be coupled with the \emph{relational} domains, that is, those that can
capture the relationships between different data objects, as well as,
the more efficient but less precise, \emph{non-relational} domains.
This flexibility in choice of domains should also be dynamic since, in
order to control the precision/efficiency trade-off, the exact domain
used may have to be changed during the analysis.
\item
The aim is to have one system that can be used for teaching, for the
development of new technologies and their evaluation and for
demonstrating their application.  For teaching the basics, we require
simple robust systems that support the core language features such as
the assignment, conditional and while commands found in all imperative
languages.  On the other hand, for research and development we need
a highly modular and extensible system so that we can plug in new
technologies without redefining the whole.  Lastly, for demonstration,
we need the system to be highly efficient with a well-structured user
interface that can verify software that is written in real languages
such as C and Java.
\end{itemize}

\subsubsection*{From \textup{CLAIR} to \textup{ECLAIR}.}
The `Combined Language and Abstract Interpretation Resource'
(CLAIR, \url{http://www.cs.unipr.it/clair/})
has been initially developed for and used in a teaching context,
as a prototype system to help in the study and experimentation
with various aspects of programming language implementation.
The CLAIR system consists of the following functional components:
\begin{itemize}
\item
A \emph{parser} module.
Lexical and syntactic analyses of the source program are performed
according to the concrete grammar of the considered target language,
leading to the computation of the corresponding
\emph{abstract syntax tree} (AST),
i.e., a Prolog term.
\item
A \emph{static semantics} module.
The AST computed in the previous phase is examined so as to compute
the static type of each program fragment and perform several
context-dependent, well-formedness checks.
\item
A \emph{dynamic semantics} module.
The concrete behavior of a program is specified by adopting
the small-step style of the structural operational semantics (SOS)
approach~\cite{Plotkin04b}.
The dynamic configurations of a running instance of the program
(i.e., the states of the transition system)
are appropriately encoded in suitable Prolog terms,
which are then evaluated according to the transition systems' rules,
leading to an interpreter for the language.
\end{itemize}
The CLAIR system features two simple (though not simplistic) languages:
SFL, a functional language; and SIL, a Pascal-like imperative language.
By programming in these languages,
students are able to see at first hand how the rather dry rules of the
structural operational semantics~\cite{Plotkin04b}
do in fact deliver a working system capable
of executing non-trivial programs,
while providing a theoretical framework
for proving interesting program properties.
These pedagogic experiences hinted at how
the same cleanness and flexibility goals
that were driving the development of a teaching tool
could also be pursued in the much more challenging context
of research, leading to the development of ECLAIR (Extended CLAIR),
whose overall aim is the analysis of mainstream programming languages.
The newly targeted system, whose specification, implementation
and evaluation is work in progress, has led to both the design
of new functional modules and the restructuring of existing ones:
\begin{itemize}
\item
The parser module has been extended with several other language instances
including, e.g., the Java bytecode language and an almost complete parser
for standard C.
\item
The static semantics module has been instrumented to save the collected
type information in a program annotation database, so as to make it
readily available to later phases.
\item
The dynamic semantics module has been refactored so as to distinguish
the implementation of the concrete semantics rules
(now given in the style of the big-step semantics,
relating programs to final configurations) from the so-called
\emph{concrete memory structure} component
that provides both the concrete memory and some of
the control flow management utilities.
The concrete memory structure component is partially implemented
in the \Cplusplus\ language so as to greatly improve the efficiency of
the generated interpreter,
which is now able to execute non-toy programs
using a reasonable amount of system resources.
\item
A new \emph{abstract semantics} module has been designed so as to allow
for the computation of the abstract semantics of the program.
As for the concrete dynamic semantics, a distinction is made
between the abstract semantics rules and the specification of the
\emph{abstract memory structure} component.
This last component is intended as
a generic interface for many of the abstract domains and operators
that have been proposed for static analysis applications,
often implemented in a foreign language such as \Cplusplus.
The abstract semantics is obtained by means of a post-fixpoint computation:
the intermediate results are saved, using the program annotation database,
as annotations attached to the nodes of the AST; care is taken so as
to allow for the specification of context sensitive analyses, whereby
different annotations can be attached to the same program fragment
depending on the calling context.
\end{itemize}

\subsubsection*{Reflections on using Prolog.}
Prolog has many benefits for the implementation of a system such as
ECLAIR, primarily due to the fact that Prolog is based on Horn clauses
which are the basis of the formal specification of all the main components.
Observe that, for the parser, we have the
added advantage that the syntax can be specified by means of the
Prolog definite clause grammar rules, which ---being a notational variant
of Horn clauses--- can be executed directly \cite{PereiraW80}.
For the other three components (static semantics, dynamic semantics
and abstract semantics), the formal specification is provided in the
form of sequents which map directly to Horn clauses;
therefore each component
can be regarded as an executable form of its specification.

It follows that, for each of the supported languages,
the implementation of the formal static and concrete semantics allows us
to perform a comprehensive series of tests so as to evaluate
different aspects of its specification and hence help us build
confidence that it
does accurately match the intended semantics; note that, in the case of
a real language such as C, the test results can be compared directly with
those obtained using a standard C compiler.
Regarding the abstract semantics rules, these also are almost directly
translated to Prolog code.  However, in this case, these rules are
incomplete in the sense that they are parametric on the abstract
domain and therefore the code implementing them must be interfaced
with specialized libraries that support a variety of abstract domains.
For ECLAIR, we use the \emph{Parma Polyhedra Library} so that all the
numerical abstractions provided by the PPL such as polyhedra,
octagons, intervals and grids can be used by the analyzer
\cite{BagnaraHRZ05SCP,BagnaraHZ05FAC,BagnaraHZ06TR}.
Note that, as the PPL provides an almost complete interface to a
number of Prolog systems, the ECLAIR to PPL interface can be written
entirely in Prolog.
We have though had to find appropriate strategies to
overcome some inherent weaknesses in Prolog.
\begin{itemize}
\item
Some components of the tool require great efficiency and are better
expressed in an imperative style.  This is the case for the
concrete memory structure that implements, among other things,
destructive updates.
As already explained, by partially implementing this structure in the
\Cplusplus\ language, we have been able to significantly improve the
performance of the interpreter.
Also, for the analyzer, it is becoming clear that coding the abstract
memory structure in Prolog (using the Prolog interface of the PPL
for the numerical abstract domains) is cumbersome.
We will thus realize that component mainly in \Cplusplus\ and
interface it to the Prolog analyzer using the generic,
portable Prolog-\Cplusplus\ interface distributed with the PPL
(see \cite{BagnaraC02TRa,BagnaraHZ06TR}).
The \Cplusplus\ module of the abstract memory structure will in turn
call Prolog for all the symbolic manipulation tasks.
\item
The lack of types in Prolog makes simple type errors hard to detect.
One possible solution is to use optional type declarations such as those
provided by the Ciao Prolog system~\cite{BuenoCCHL-GP97}.
However, such non-standard annotations, could lead to a loss of portability
of the ECLAIR system.
A possibility we are considering is the use of the TCLP prescriptive type
system \cite{FagesC01}, the main problem being that it does not currently
support SWI-Prolog.
\item
Well-known deficiencies of Prolog module systems (first of all their flat,
not hierarchical nature) make the development of a modular system like
ECLAIR more difficult than it ought to be.
\end{itemize}
\subsubsection*{Conclusion with ongoing and future work.}
We have sketched the overall structure of ECLAIR, outlining the main
methodologies and technologies underlying its components.  We also
have justified why we chose Prolog as the implementation language.
Apart from the continuing development of the concrete and abstract
interpreters, current work includes investigating several applications
for the analyzer, particularly: string cleanness, absence of integer
overflows, correctness of array operations, inference of ranking
functions and termination analysis.  Further interesting extensions of
the system are foreseen. In particular, the development of
\emph{compiler} modules (including code generation and optimizations):
as well as the obvious applications in a teaching context, this will
allow to evaluate the usefulness of the information extracted by the
abstract semantics module also from the standpoint of optimized
compilation.


\begin{thebibliography}{10}

\bibitem{BagnaraC02TRa}
R.~Bagnara and M.~Carro.
\newblock Foreign language interfaces for {Prolog}: A terse survey.
\newblock Quaderno 283, Dipartimento di Matematica, Universit\`a di Parma,
  Italy, 2002.
\newblock Version 1. Available at \url{http://www.cs.unipr.it/Publications/}.

\bibitem{BagnaraHPZ07TR}
R.~Bagnara, P.~M. Hill, A.~Pescetti, and E.~Zaffanella.
\newblock On the design of generic static analyzers for modern imperative
  languages.
\newblock Technical Report~{\tt arXiv:cs.PL/0703116}, Dipartimento di
  Matematica, Universit\`a di Parma, Italy, 2007.
\newblock Available from \url{http://arxiv.org/}.

\bibitem{BagnaraHRZ05SCP}
R.~Bagnara, P.~M. Hill, E.~Ricci, and E.~Zaffanella.
\newblock Precise widening operators for convex polyhedra.
\newblock {\em Science of Computer Programming}, 58(1--2):28--56, 2005.

\bibitem{BagnaraHZ05FAC}
R.~Bagnara, P.~M. Hill, and E.~Zaffanella.
\newblock Not necessarily closed convex polyhedra and the double description
  method.
\newblock {\em Formal Aspects of Computing}, 17(2):222--257, 2005.

\bibitem{BagnaraHZ06TR}
R.~Bagnara, P.~M. Hill, and E.~Zaffanella.
\newblock The {Parma Polyhedra Library}: Toward a complete set of numerical
  abstractions for the analysis and verification of hardware and software
  systems.
\newblock Quaderno 457, Dipartimento di Matematica, Universit\`a di Parma,
  Italy, 2006.
\newblock Available at \url{http://www.cs.unipr.it/Publications/}. Also
  published as {\tt arXiv:cs.MS/0612085}, available from
  \url{http://arxiv.org/}.

\bibitem{BuenoCCHL-GP97}
F.~Bueno, D.~Cabeza, M.~Carro, M.~Hermenegildo, P.~L\'{o}pez-Garc\'{\i}a, and
  G.~Puebla.
\newblock The {Ciao Prolog} system. {Reference} manual.
\newblock Technical Report {CLIP}3/97.1, School of Computer Science, Technical
  University of Madrid (UPM), 1997.
\newblock Available from http://www.clip.dia.fi.upm.es/.

\bibitem{Cousot05b}
P.~Cousot.
\newblock The verification grand challenge and abstract interpretation.
\newblock In {\em Verified Software: Theories, Tools, Experiments (VSTTE)}, ETH
  {Z\"urich}, Switzerland, 2005.
\newblock Position paper.

\bibitem{CousotC77}
P.~Cousot and R.~Cousot.
\newblock Abstract interpretation: A unified lattice model for static analysis
  of programs by construction or approximation of fixpoints.
\newblock In {\em Proceedings of the Fourth Annual ACM Symposium on Principles
  of Programming Languages}, pages 238--252, New York, 1977. ACM Press.

\bibitem{CousotC92fr}
P.~Cousot and R.~Cousot.
\newblock Abstract interpretation frameworks.
\newblock {\em Journal of Logic and Computation}, 2(4):511--547, 1992.

\bibitem{CousotC92}
P.~Cousot and R.~Cousot.
\newblock Inductive definitions, semantics and abstract interpretation.
\newblock In {\em Proceedings of the Nineteenth Annual ACM Symposium on
  Principles of Programming Languages}, pages 83--94, Albuquerque, New Mexico,
  USA, 1992. ACM Press.

\bibitem{CousotCFMMMR05}
P.~Cousot, R.~Cousot, J.~Feret, L.~Mauborgne, A.~Min\'e, D.~Monniaux, and
  X.~Rival.
\newblock The {ASTR\'EE} analyzer.
\newblock In M.~Sagiv, editor, {\em Programming Languages and Systems,
  Proceedings of the 14th European Symposium on Programming}, volume 3444 of
  {\em Lecture Notes in Computer Science}, pages 21--30, Edinburgh, UK, 2005.
  Springer-Verlag, Berlin.

\bibitem{FagesC01}
F.~Fages and E.~Coquery.
\newblock Typing constraint logic programs.
\newblock {\em Theory and Practice of Logic Programming}, 1(6):751--777, 2001.

\bibitem{Kahn87}
G.~Kahn.
\newblock Natural semantics.
\newblock In F.-J. Brandenburg, G.~Vidal-Naquet, and M.~Wirsing, editors, {\em
  Proceedings of the 4th Annual Symposium on Theoretical Aspects of Computer
  Science}, volume 247 of {\em Lecture Notes in Computer Science}, pages
  22--39, Passau, Germany, 1987. Springer-Verlag, Berlin.

\bibitem{PereiraW80}
F.~C.~N. Pereira and D.~H.~D. Warren.
\newblock Definite clause grammars for language analysis - a survey of the
  formalism and a comparison with augmented transition networks.
\newblock {\em Artificial Intelligence}, 13(3):231--278, 1980.

\bibitem{Plotkin04b}
G.~D. Plotkin.
\newblock A structural approach to operational semantics.
\newblock {\em Journal of Logic and Algebraic Programming}, 60--61:17--139,
  2004.

\bibitem{Schmidt95}
D.~A. Schmidt.
\newblock Natural-semantics-based abstract interpretation (preliminary
  version).
\newblock In A.~Mycroft, editor, {\em Static Analysis: Proceedings of the 2nd
  International Symposium}, volume 983 of {\em Lecture Notes in Computer
  Science}, pages 1--18, Glasgow, UK, 1995. Springer-Verlag, Berlin.

\bibitem{Schmidt97}
D.~A. Schmidt.
\newblock Abstract interpretation of small-step semantics.
\newblock In M.~Dam, editor, {\em Analysis and Verification of Multiple-Agent
  Languages}, volume 1192 of {\em Lecture Notes in Computer Science}, pages
  76--99. Springer-Verlag, Berlin, 1997.
\newblock 5th LOMAPS Workshop Stockholm, Sweden, June 24--26, 1996, Selected
  Papers.

\bibitem{Schmidt98}
D.~A. Schmidt.
\newblock Trace-based abstract interpretation of operational semantics.
\newblock {\em {LISP} and Symbolic Computation}, 10(3):237--271, 1998.

\end{thebibliography}
\newcommand{\noopsort}[1]{}\hyphenation{ Ba-gna-ra Bie-li-ko-va Bruy-noo-ghe
  Common-Loops DeMich-iel Dober-kat Di-par-ti-men-to Er-vier Fa-la-schi
  Fell-eisen Gam-ma Gem-Stone Glan-ville Gold-in Goos-sens Graph-Trace
  Grim-shaw Her-men-e-gil-do Hoeks-ma Hor-o-witz Kam-i-ko Kenn-e-dy Kess-ler
  Lisp-edit Lu-ba-chev-sky Ma-te-ma-ti-ca Nich-o-las Obern-dorf Ohsen-doth
  Par-log Para-sight Pega-Sys Pren-tice Pu-ru-sho-tha-man Ra-guid-eau Rich-ard
  Roe-ver Ros-en-krantz Ru-dolph SIG-OA SIG-PLAN SIG-SOFT SMALL-TALK Schee-vel
  Schlotz-hauer Schwartz-bach Sieg-fried Small-talk Spring-er Stroh-meier
  Thing-Lab Zhong-xiu Zac-ca-gni-ni Zaf-fa-nel-la Zo-lo }

\end{document}